\documentclass[12pt]{article}                              

\usepackage[all]{xy}

\usepackage{amsmath,amsthm,amsfonts,amscd,amssymb,eucal,makeidx}
\numberwithin{equation}{section}
 
\def\A{{\cal A}}
\def\B{{\cal B}}

\def\D{{\cal D}}

\def\H{{\cal H}}

\def\K{{\cal K}}
\def\L{{\cal L}}

\def\O{{\cal O}}

\def\P{{\cal P}}

\def\T{{\cal T}}

\def\Tr{{\mathrm {Tr}}}

\def\b{\beta}
        
        \def\D{\Delta}

\def\l{\lambda}       \def\L{\Lambda}

\def\PSL{{{\rm PSL}(2,\mathbb R)}}
\def\SL{{{\rm SL}(2,\mathbb R)}}

\def\Mob{G}

\def\Iu{I_{\cap}}
\def\Ir{I_{\supset}}

\newtheorem{Thm}{Theorem}[section]
\newtheorem{Cor}[Thm]{Corollary}
\newtheorem{Prop}[Thm]{Proposition}
\newtheorem{Lemma}[Thm]{Lemma}
\theoremstyle{definition}

\theoremstyle{remark}

 \makeindex

\begin{document}

\title{\bf Nuclearity and Thermal States
\\in Conformal Field Theory}
\author{{\sc Detlev Buchholz}\footnote{buchholz@theorie.physik.uni-goettingen.de, dantoni@mat.uniroma2.it,\,\nobreak longo@mat.uniroma2.it}\\[3mm] 
Institut f\"ur Theoretische Physik, Universit\"at G\"ottingen,
\\ 37077 G\"ottingen, Germany\\[4mm]
{\sc Claudio D'Antoni$^*$\footnote{Supported by MIUR, GNAMPA-INDAM and EU network ``Quantum Spaces - Noncommutative Geometry" 
HPRN-CT-2002-00280
}  {\small and} \sc Roberto Longo $^{*\dag}$}\\[3mm] 
Dipartimento di Matematica, Universit\`a di Roma ``Tor
Vergata'', \\ Via della Ricerca Scientifica, 1 - I-00133 Roma, Italy}

\date{}
\maketitle
\centerline{\sl Dedicated to L\'aszl\'o Zsid\'o on the occasion of his sixtieth birthday}


\begin{abstract} 
We introduce a new type of spectral density condition, that we call $L^2$-nuclearity. One formulation concerns lowest weight unitary representations of $SL(2,\mathbb R)$ and turns out to be equivalent to the existence of characters. A second formulation concerns inclusions of local observable von Neumann algebras in Quantum Field Theory. 
We show the two formulations to agree in chiral Conformal QFT and, starting from the trace class condition $\Tr(e^{-\b L_0})<\infty$ for the conformal Hamiltonian $L_0$, we infer and naturally estimate the Buchholz-Wichmann nuclearity condition and the (distal) split property.
As a corollary, if $L_0$ is log-elliptic, the Buchholz-Junglas set up is realized and so there exists a $\beta$-KMS state for the translation dynamics on the net of C$^*$-algebras for every inverse temperature $\beta>0$.
We include further discussions on higher dimensional spacetimes. In particular, we verify that $L^2$-nuclearity is satisfied for the scalar, massless Klein-Gordon field.
\end{abstract}

\section{Introduction}

As is known the general principles of Quantum Field Theory (locality, Poin\-car\'e covariance, positivity of the energy) are compatible with physically unrealistic models. A simple example is obtained by the infinite tensor product of copies of a fixed model: although the basic structure is preserved under this infinite tensoring, the degrees of freedom may arbitrarily increase. 

In order to select physically acceptable theories, two kind of conditions have been introduced in particular. The first are compactness/nuclearity conditions. The {\em Haag-Swieca compactness} condition requires the set of vectors localized in a given bounded spacetime region with a common energy bound to be a (norm) pre-compact subset of the Hilbert space.

A quantitative, efficient refinement of the above compactness is the {\em Buchholz-Wichmann nuclearity condition} \cite{BW}: for given double cone $\O$ of the Minkowski spacetime and $\b > 0$ the map
\[
\Phi_{\O}(\b): \A(\O)\to \H,\quad X\mapsto e^{-\b H}X\Omega
\]
has to be a nuclear operator. Here $\H$ is the state Hilbert space, $H$ is the energy operator and $\A(\O)$ is the von Neumann algebra of observables localized within $\O$. Possibly conditions on the nuclear norm $||\Phi_{\O}(\b)||_1$ as $\O$ increases or $\b\to 0$ may be added. There are several different versions of the nuclearity condition that reflect structural properties of the considered theory.

The second type condition is the {\em split property}. This property is intrinsically encoded in the net $\A$ of local observable algebras. It has several equivalent formulations; we shall state here that $\A$ is defined to be split if the natural map
\[
X\otimes X' \in\A(\O) \odot\A(\tilde\O)' \mapsto X X' \in\A(\O)\cdot \A(\tilde\O)'
\]
(from the algebraic tensor product) extends to an isomorphism between the von Neumann algebras $\A(\O)\otimes\A(\tilde\O)'$ and $\A(\O)\vee\A(\tilde\O)'$ (statistical independence). Here $\O\subset\tilde\O$ are double cones and $\O\Subset\tilde\O$. \footnote{$\O\Subset\tilde\O$ means that the closure of $\O$ is contained in the interior of $\tilde\O$.}

Several structure properties follow from the split assumption, for example one can derive the local current algebra (in integrated form) \cite{DL}. The original motivation to introduce the nuclearity was indeed to derive the split property \cite{BW}. The nuclearity condition implies the split property, is directly analyzable and has a number of important consequences in itself, as the construction of thermal equilibrium states.

This paper primarily concerns conformal quantum field theory, say chiral conformal QFT on the circle. In this context the energy operator is usually replaced with the conformal Hamiltonian $L_0$ and so there is a another natural nuclearity condition, namely the {\em trace class condition} for the semigroup generated by $L_0$:
\[
\Tr(e^{-\b L_0}) <\infty,\quad \b >0\ ,
\]
namely the characters have to be defined or, from the Physical viewpoint, Gibbs states for $L_0$ must exist at inverse temperature $\b>0$.

It is known that the trace class condition for all $\b>0$ implies the split property, although the proof is rather indirect \cite{DLR}. It is natural to study the relations between the trace class condition and the more physical Buchholz-Wichmann nuclearity property; in particular one would derive this latter property from the trace class condition which is a standard assumption in most approaches to conformal QFT. Note the different character of the two conditions: the trace class condition only refers to the associated unitary representation of the M\"obius group while Buchholz-Wichmann nuclearity is expressed in terms of local observables too.

We shall indeed prove that Buchholz-Wichmann nuclearity follows from the trace class condition, with a rather precise description of this dependence.

One most interesting ingredient of our analysis is the introduction of a new nuclearity condition that we call {\em $L^2$-nuclearity condition}. The key point is that this condition can be phrased in two different ways: one only refers to the unitary representation $U$ of the M\"obius group, the other to the net of local algebras $\A$, so we will have a bridge to relate the trace class condition to the Buchholz-Wichmann nuclearity. To be explicit denote by $K_I\equiv -\frac{1}{2\pi}\log\D_I$ the generator of the dilation one-parameter unitary group associated with the interval $I$, namely eq. \eqref{dil1} below holds. Then the $L^2$-nuclearity condition requires that the operator\footnote{We often denote by the same symbol a linear operator and its closure.}
\[
T^U_{\tilde I, I} \equiv \D^{1/4}_{\tilde I}\D^{-1/4}_I
\]
be a trace class operator on $\H$. Clearly this is a condition on the representation $U$.

Now the operator $\D_I$ is the Tomita-Takesaki modular operator associated with the pair $\A(I)$ and the vacuum vector, so the $L^2$-nuclearity condition $||\D^{1/4}_{\tilde I}\D^{-1/4}_I ||_1 <\infty$ can be expressed in term of the net $\A$. It is then straightforward to infer {\em modular nuclearity} from $L^2$-nuclearity. Modular nuclearity \cite{BDL1} is a sort of ``local version" of Buchholz-Wichmann nuclearity where the energy operator is replaced with a modular operator (generator of a ``local dynamics''). As shown in \cite{BDL2} modular nuclearity and Buchholz-Wichmann nuclearity are closely related conditions, indeed they are equivalent with the assumption of certain growth conditions.

Our chain of implications becomes closed by showing that $L^{2}$-nuclearity is equivalent to the trace class condition. Indeed we shall have the equality
\begin{equation}\label{l2tr}
||T^U_{\tilde I,I}||_1 = \Tr(q^{L_0}) \ ,
\end{equation}
where $\tilde I$ is a symmetric interval of $\mathbb R$ and $I = q\tilde I$ is a dilated interval, $0<q<1$.
As we shall see that $||T^U_{\tilde I,I}||_1 <\infty$ immediately implies that $\A(I)\subset\A(\tilde I)$ is split, we shall have in this way a precise dependence from the trace class condition at a fixed $\b>0$ to the split property at a certain distance (distal split property).

This paper is organized as follows. We begin by recalling a formula long pointed out by Schroer and Wiesbrock (see \cite{S}), that we prove here. This formula concerns an unbounded similarity between the semigroup generated by the conformal Hamiltonian and a dilation one-parameter unitary group. 

We then discuss certain operator identities that are quite powerful for our analysis and have their own interest. The first
\[
T_{\tilde I, I}T^*_{\tilde I, I} =  q^{2L_0},
\]
directly relates the conformal Hamiltonian and the $T$ operators. The second one
\[
e^{-2sL_0} = e^{-\tanh(\frac{s}{2})H}e^{-\sinh(s)H'}e^{-\tanh(\frac{s}{2})H}
\]
relates the conformal Hamiltonian with the translation Hamiltonian $H$ and its ray inversion conjugate $H'$. This shows that $T$ operators are naturally related both to the translation semigroup (see also Proposition \ref{fin2}) and to the rotation semigroup and are thus a natural link between these semigroup.

We now come to one most interesting consequence of our analysis, the existence of temperature states for the translation dynamics, starting from natural requirements for the conformal Hamiltonian, that partly motivated our paper.

Quantum Field Theory is a scheme to analyze finitely many quantum particles and one primarily considers the homogeneous fundamental state, the vacuum state; at this stage there is no thermodynamical consideration as the vacuum is state of zero temperature. In recent years however the study of finite-temperature states has acquainted a definite interest because of various reasons, in particular the desire to study extreme situations where gravitational effects make matter so dense that system becomes thermodynamical (e.g. black holes). 

A general construction of thermal states in Quantum Field Theory has been made by Buchholz and Junglas \cite{BJ}. One of our aim is to derive their assumptions in the context of conformal QFT starting from natural assumptions in this context.

As said, with our formulas we can  easily discuss the relations among various forms of nuclearity. As a corollary we see that if $L_0$ is log-elliptic in the sense that (cf. \cite{KL})
\[
\log\Tr(e^{-sL_0}) \sim {\rm const.}\frac{1}{s^{\alpha}}, \qquad s\to 0^+,
\]
for some $\alpha >0$, then the Buchholz-Junglas nuclearity assumptions are satisfied and so there exists a $\beta$-KMS state with respect to translations on the net of C$^*$-algebras for every inverse temperature $\beta>0$ (see also \cite{L} for general considerations).

Note that log-ellipticity naturally holds with $\alpha=1$ in rational conformal field theory (modular nets) \cite{KL}.

We end our paper with a discussion concerning QFT on higher dimensional space-times.
In particular we will use our results for chiral conformal QFT to verify the $L^2$ nuclearity condition for the neutral, massless, free field on the Minkowski spacetime.

\section{An operator identity associated with $\SL$}

The upper and the right semicircle will be denoted respectively by $\Iu$ and $\Ir$. We shall often pass from the ``circle picture" to the ``real line picture", namely we identify $S^1\setminus \{-1\}$ with $\mathbb R$ by the stereographic map. Then $\Iu \simeq (0,\infty)$ and $\Ir \simeq (-1,1)$. We shall denote by $G$ the universal cover of M\"obius group ($\simeq\PSL$), that acts on $S^1$ as usual. With $I$ an interval of $S^1$, we denote by $\L_I$ the one-parameter subgroup of $\PSL$ of  ``dilations'' associated with $I$; namely 
\[
\L_{\mathbb R^+}(s): x\in\mathbb R\mapsto e^s x\in\mathbb R\ .
\]
and $\L_I$ is then defined by conjugation by any $g\in \PSL$ such that $g\Iu=I$. 

$\L_I$ has a unique lift to a one-parameter subgroup of $G$, that we will still denote by $\L_I$.

Given a unitary, positive energy representation $U$ of $G$ on a 
Hilbert space $\H$, we shall denote by $L_0\equiv L^U_{0}$ the 
conformal Hamiltonian, namely the infinitesimal generator of the 
rotation subgroup. Thus $L_0$ is a positive selfadjoint operator on 
$\H$. 
We shall denote by $K_I$ the selfadjoint generator on $\H$ of $U(\L_I(\cdot))$, namely
\begin{equation}
e^{isK_I} = U(\L_I(s)) \ .
\end{equation}
We shall set $K_I\equiv -\frac{1}{2\pi}\log\D_I$, thus $\D_I \equiv \D_{I,U}$ is the unique positive, non-singular, selfadjoint operator on $\H$ such that
\begin{equation}\label{dil1}
\D_I^{is} = U(\L_I(-2\pi s))\ .
\end{equation}
Given the positive energy, unitary representation $U$ of $G$ on $\H$ we choose an anti-unitary involution $J=J_{\Iu}$ such that
\[
JU(g)J = U(rgr^{-1})
\]
where $r$ is any pre-image in $G$ of the M\"obius transformation $z\mapsto\bar z$; then one defines $J_I$ for any interval $I$ by requiring that $J_{gI}\equiv U(g)J_{I}U(g)^*$. Following \cite{BGL3} we define the unbounded anti-linear involution $S_I\equiv J_I\D_I^{1/2}$ and set
\[
H(I)\equiv \{\xi\in\H: S_I\xi=\xi\}\ .
\]
$H(I)$ is a real, standard Hilbert subspace of $\H$ and one has (see \cite{BGL3}):
\begin{Prop}
    $\D_I$ is the modular operator associated with $H(I)$.
\end{Prop}
Clearly the family of real Hilbert spaces $H(I)$ is covariant for the representation $U$. A non-trivial fact is that if $U$ is a representation of $SL(2,\mathbb R)$ (not of its cover), then $\{H(I)\}$ is isotone, namely
\[
I\subset\tilde I\Rightarrow  H(I)\subset H(\tilde I) \ .
\]
Given an inclusion of real, standard Hilbert subspaces $H\subset\tilde H$ of $\H$ we shall consider the linear operator on $\H$
\[
T_{\tilde H, H}(\l)\equiv \D_{\tilde H}^{\l}\D_{H}^{-\l},\quad 0\leq\Re \l\leq1/2\ .
\]
For each $\l$, $T_{\tilde H, H}(\l)$ is densely defined, bounded and $||T_{\tilde H, H}(\l)||\leq 1$. Denoting the closure by the same symbol $T_{\tilde H, H}(\l)$, the map $\l\mapsto T_{\tilde H, H}(\l)$ is holomorphic in the strip $0<\Re \l<1/2$ and continuous on the closure. These facts can be proved by the same arguments as in \cite{BDL1}. 

Given the unitary representation $U$ of $G$ as above, we then set
\[
T_{\tilde I, I}(\l)\equiv T_{H(\tilde I), H(I)}(\l)
\]
so we have:
\begin{Cor}\label{hol}
If $U$ is a positive energy, unitary representation of $SL(2,\mathbb R)$ on $\H$,
and $I\subset\tilde I$ are intervals of $S^1$, then the associated operator
$T_{\tilde I, I}(\l)$
is bounded with $||T_{\tilde I, I}(\l)||\leq 1$ and the map $\l\mapsto T_{\tilde I, I}(\l)$ is holomorphic in the strip $0<\Re \l<1/2$ and continuous on its closure.
\end{Cor}
We shall see that Cor. \ref{hol} holds also for all positive energy, unitary representations of $G$, but the proof is non-trivial as $H(I)$ is not included in $H(\tilde I)$ in this general case.

The case $\l=1/4$ is of particular relevance and we set
\[
T_{\tilde I, I}\equiv T_{\tilde I, I}(1/4)\ .
\]
We now prove a formula pointed out by Schroer and Wiesbrock \cite{S}. We give here below a proof in the case $U$ is a representation of $SL(2,\mathbb R)$, a case that covers most needs in this paper. The proof will be continued in the Appendix \ref{Am} to treat the general case of representations of the cover $G$.
\begin{Thm} {\rm cf. \cite{S}.}
\label{magic'} 
Let $U$ be a positive energy unitary representation of 
$G$. For every $s\geq 0$, the following identity holds
\begin{equation}\label{magic} 
    \D_{1}^{1/4}\D_{2}^{-is}\D_{1}^{- 1/4} = e^{-2\pi s L_0}
\end{equation}
where $\D_{1}\equiv\D_{I_{\cap}}$, $\D_{2}\equiv\D_{I_{\supset}}$ and $L_0$ are associated with $U$.

More precisely the domain of $\D_{1}^{1/4}\D_{2}^{-is}\D_{1}^{- 1/4}$ is a core
for $\D_{1}^{- 1/4}$ and the closure of $\D_{1}^{1/4}\D_{2}^{-is}\D_{1}^{- 1/4}$ is equal to $e^{-2\pi s L_0}$.
\end{Thm}
\proof We split the proof in parts; a part may rely on a (temporary) independent extra assumption.

{\it Part 1.}
We denote by $k_1 , k_2 $ and $l_0$ the elements in the Lie algebra $sl(2,\mathbb C)$ corresponding to $K_1\equiv  K_{I_1}$, $K_2\equiv  K_{I_2}$ and $L_0$. Then
\[
[k_1 , k_2] = -il_0, \quad [k_1, l_0] = k_2\ .
\]
Denoting by Ad$(g)$ the adjoint action of $g\in G$ on $sl(2,\mathbb C)$ we then have
\[
{\rm Ad}(e^{-2\pi i t k_1})(k_2) = \sum_{n=0}^{\infty}\frac{t^n}{n!}\delta_{k_1}^n(k_2) = \cosh(2\pi t)k_2 - \sinh(2\pi t)l_0
\]
where $\delta_{k_1}\equiv 2\pi[k_1,\,\cdot\,]$, therefore for all $s,t\in\mathbb R$ we have the identity in $G$
\begin{equation*}
e^{-2\pi i t k_1}e^{2\pi i s k_2}e^{2\pi i t k_1}
=e^{2\pi i s \big(\cosh(2\pi t)k_2 - \sinh(2\pi t)l_0\big)}
\end{equation*}
that of course gives the operator identity
\begin{equation}\label{BCH}
e^{-2\pi i t K_1}e^{2\pi i s K_2}e^{2\pi i t K_1}
=e^{2\pi i s \big(\cosh(2\pi t)K_2 - \sinh(2\pi t)L_0\big)}
\end{equation}
Consider the right hand side of eq. \eqref{BCH} that we denote by $W_s(t)$. Given $r>1/2$, by Lemma \ref{NW} 
there exist $s_0>0$, a dense set $\cal D\subset\H$ of joint analytic vectors for $K_1, K_2, L_0$ such that, for any fixed $s\in \mathbb R$ with $|s|\leq s_0$ and $\eta\in\cal D$,
the vector-valued function
\[
t\mapsto W_s(t)\eta \equiv e^{2\pi i s \big(\cosh(2\pi t)K_2 - \sinh(2\pi t)L_0\big)}\eta
\]
has a bounded analytic continuation in the ball $B_r\equiv\{z\in \mathbb C: |z|<r\}$.

Consider now the left hand side of eq. \eqref{BCH}. By definition it is equal to $\D_1^{it}\D_2^{-is}\D_1^{-it}$.

{\it Part 2.} 
In this part we assume that $U$ is a representation of $SL(2,\mathbb R)$ (rather than of its cover $G$). 

Now the map
\begin{equation}\label{an2}
t\in\mathbb R\mapsto \D_1^{it}\D_2^{-is}\D_1^{-it}
\end{equation}
has a uniformly bounded, strongly operator continuous, analytic extension in the strip $S(0,-1/2)\equiv\{z\in\mathbb C: -1/2<\Im z < 0\}$. Indeed $\D_1^{\l}\D_2^{-is}\D_1^{-\l}=T_{I_1,I_{1,s}}(\l)\D_2^{-is}$ where $I_{1,s}=\L_{I_2}(2\pi s)I_1\subset I_1$, so the analyticity of \eqref{an2} follows from Lemma \ref{hol} in the $SL(2,\mathbb R)$-case. 

Taking matrix elements
\[
(\eta, \D_1^{it}\D_2^{-is}\D_1^{-it}\xi) = (W_s(t)^*\eta,\xi) =(W_{-s}(t)\eta,\xi)
\]
with $\eta\in\cal D$ and $\xi$ an entire vector for $\D_1$, the both functions defined by left anf right side of the above equation have an analytic extension in $S(0,-1/2)\cap B_r$.

Taking the value at $t=-i/4$ we have 
\[
(\eta, \D_1^{1/4}\D_2^{-is}\D_1^{-1/4}\xi) = (e^{-2\pi sL_0}\eta,\xi)=  (\eta,e^{-2\pi sL_0}\xi)
\]
hence the closure of $\D_1^{1/4}\D_2^{-is}\D_1^{-1/4}$ is equal to $e^{-2\pi sL_0}$ if $s$ is real and $|s|\leq s_0$, hence for all $s\in\mathbb R$ by the group property. 

This ends the proof in the $SL(2,\mathbb R)$-case.
\endproof
The proof is continued in Appendix \ref{Am}.
We now recall the following result by Nelson used in Th. \ref{magic'}.
\begin{Lemma} {\rm \cite{NW}}\label{NW}
Let $U$ be a unitary representation of a Lie group $\cal G$ on a Hilbert space $\H$ and $X_1,X_2, \dots X_n$ a basis for the associated Lie algebra generators. There exist a neighborhood $\cal U$ of the origin in $\mathbb C^n$ and a dense set of vectors $\cal D\subset\H$ of smooth vectors for $U$ such that 
\[
\sum_{k=0}^{\infty}\frac{||(u_1 X_1 + u_2 X_2+\cdots + u_n X_n)^k\eta||}{k!} <\infty
\]
for all $(u_1,u_2,\dots u_n)\in\cal U$ and $\eta\in\cal D$.
\end{Lemma}

\section{Trace class property and $L^2$-nuclearity for representations of $\SL$}

\subsection{First formula}
\label{T11}

Given intervals $I\Subset \tilde I$ we consider the {\em inner distance} 
between $\tilde I$ and $I$ to be the number $\ell(\tilde I, I)$ 
defined as follows. First assume $\tilde I = \Iu$ and $I$ symmetric with respect to the vertical axis, namely $I$ has boundary points $\{z, -\bar z\}$ with $z\in\Iu\cap\Ir$. Then $z = \L_{\Ir}(s)1$ for a unique $s>0$ and we put $\ell(\tilde I, I)\equiv s$. For a general inclusion $I\Subset \tilde I$ there exists a unique $g\in G$ such that $g\tilde I = \Iu$ and $gI = -\overline{gI}$ as above and we put $\ell(\tilde I, I)\equiv \ell(g\tilde I, gI)$. Of course $\ell(\tilde I, I)$ can be any positive real number. A simple expression is given in the real line picture if $I,\tilde I\subset \mathbb R$ are symmetric intervals: then $\ell(\tilde I, I)=\log(d_{\tilde I}/{d_I})$ where $d_I$ is the usual length of $I$.
(See Appendix \ref{app} for more).

If $A\in B(\H)$, the nuclear norm $||A||_1$ of $A$ is the $L^1$ norm, namely 
$||A||_1 \equiv \Tr(|A|)$ where $|A|\equiv \sqrt{A^* A}$; the Hilbert-Schmidt norm is given by $||A||_2=\Tr(A^*A)^{1/2}$. 

We shall consider the property that $||T^{U}_{\tilde I, I}||_1  
<\infty$, that we call {\em $L^2$-nuclearity} (with respect to 
$I\Subset\tilde I$). Note that $||T^{U}_{\tilde I, I}||_1 $
depends only on $\ell(\tilde I, I)$, namely $||T^{U}_{\tilde I, I}||_1 $ 
does not change if we replace $I\Subset\tilde I$ by 
$hI\Subset h\tilde I$ with $h\in G$.

\begin{Prop} \label{fin}
In every positive energy unitary representation $U$, we have
\[
T^{U}_{\tilde I, I} =  e^{-sL_0}\D_{2}^{is/2\pi }
\]   
where $\tilde I = \Iu$, $I\Subset\tilde I$ is symmetric w.r.t. the vertical axis, $s=\ell(\tilde I, I)$ and $\D_2$ is as above. Therefore
\begin{equation}\label{TT}
||T^{U}_{\tilde I, I}||_1  =  ||e^{-sL_0}||_1
\end{equation}
for any inclusion $I\Subset\tilde I$ such that $s=\ell(\tilde I, I)$.
\end{Prop}
\proof 
Denote by $\{z,-\bar z\}$ the  boundary points of $I$.  By multiplying both sides of formula \eqref{magic} by
$\D_{2}^{is}$ on the right, we get the equality
 \begin{align*}
e^{-2\pi s L_0}\D_{2}^{is }= &  \D_{1}^{1/4}\big(\D_{2}^{-is}\D_{1}^{- 1/4}\D_{2}^{is}\big) \\
= &  \D_{1}^{1/4}\big(U(\L_{I_2}(2\pi s))\D_{1}^{- 1/4}U(\L_{I_2}(-2\pi s))\big) \\
= &  \D_{1}^{1/4}\D_{I_{1,2\pi s}}^{- 1/4}
\end{align*}
where $I_{1,s}\equiv \L_{I_2}(s)I_1$; that is to say  $T^{U}_{\tilde I, I} = e^{- s L_0}\D_{2}^{is/2\pi }$
where  $s=\ell(\tilde I, I)$. Since then $\D_{2}^{is/2\pi }$ is unitary we are done.
\endproof
As a consequence we have a key equation for the $T$ operator
\begin{equation}\label{T1}
T_{\tilde I, I}T^*_{\tilde I, I} =  e^{-2sL_0},\quad s\equiv \ell(\tilde I, I)
\end{equation}
with $\tilde I$ the upper semicircle and $I$ symmetric w.r.t. the vertical axis.
Of course by M\"obius covariance we have a general formulation of the above proposition and the above equation for arbitrary inclusions $I\Subset\tilde I$. In particular formula \eqref{TT} holds true for any inclusion of intervals $I\Subset\tilde I$ with $s=\ell(\tilde I, I)$.

\subsection{Second formula}
\label{T3}

Denote by $\tau$ the one parameter group of translations on $\mathbb R$ and $\tau'$ the translations associate with $(-\infty,0)$, namely the conjugate group of $\tau$ by the ray inversion map $x\to -1/x$. Both $\tau_a$ and $\tau'_{-a'}$ maps $\mathbb R_+$ into itself for positive $a$ and $a'$.

Set $I\equiv \mathbb R_+$ and $I_{a',a}\equiv\tau'_{-a'}\tau_a I$ with $a,a'>0$, so that $I_{a',a}\Subset I$. Let $U$ be a positive energy unitary representation of $\Mob$ and denote by $H\equiv H_I$ and $H'\equiv H'_I = H_{I'}$ the positive generators of the one-parameter unitary subgroups corresponding to $\tau$ and $\tau'$. We have:
\begin{Prop}\label{fin2}
\[
T^U_{I,I_{a',a}}= e^{-a' H'_I}e^{-a H_I}e^{-ia H_I}e^{ia' H'_I}\ .
\]
\end{Prop}
\proof
Indeed
\begin{equation}\label{fft}
\begin{aligned}
T^U_{I,I_{a',a}} &\equiv \D_I^{1/4}\D_{I_{a',a}}^{-1/4} \\
&= \D_I^{1/4}e^{-ia' H'_I}e^{ia H_I}\D_I^{-1/4}e^{-ia H_I}e^{ia' H'_I}\\
&=\D_I^{1/4}e^{-ia' H'_I}\D_I^{-1/4}\D_I^{1/4}e^{ia H_I}\D_I^{-1/4}e^{-ia H_I}e^{ia' H'_I}\\
&= e^{-a'H'_I}e^{-aH_I}e^{-ia H_I}e^{ia' H'_I}
\end{aligned}
\end{equation}
where we have used the Borchers commutation relation
\[
\D_I^{is}e^{ia H_I}\D_I^{-is} = e^{i(e^{-2\pi s}) aH_I}
\]
and the analogous one with $H'$ instead of $H$. If $a>0$, the above equation holds true for all complex $s$ with $-1/2\geq\Im s \geq 0$ and we have applied it with $s = -i/4$  \cite{B}.
\endproof
As a consequence we have another key equation for the $T$ operator:
\begin{equation}\label{T2}
T_{I, I_{a',a}}T^*_{I, I_{a',a}} = e^{-aH_I}e^{-2a'H'_I}e^{-aH_I}\ .
\end{equation}
Note also that by Prop. \ref{fin} we also have
\[
||T_{I,I_{a',a}}||_1 = \Tr(e^{-\ell(I,I_{a',a})L_0}) = \Tr(e^{-2\sinh^{-1}(\ell'(I,I_{a',a}))L_0}) = \Tr(e^{-2\sinh^{-1}(\sqrt{aa'})L_0})
\]
where  $\ell'(I,I_{a',a})=\sqrt{a'a}$ is the {\em second inner distance} (Appendix \ref{app}), thus $\ell'=\sinh(\frac{\ell}{2})$ by Prop. \ref{ll'}. \footnote{As $2L_0= H + H'$ one could more directly use the Golden-Thompson inequality $\Tr(e^{-aH}e^{-2a' H'}e^{-aH})\geq \Tr(e^{-2aH - 2a'H')})$ (see \cite{R}), yet this only gives an inequality.}

We now have some of our basic formulas.
\begin{Thm} In any positive energy unitary representation of $\Mob$ we have
\begin{equation}\label{m1}
e^{-2sL_0} = e^{-\tanh(\frac{s}{2})H}e^{-\sinh(s)H'}e^{-\tanh(\frac{s}{2})H}
\end{equation}
therefore
\begin{equation}\label{m2}
e^{-2sL_0} \leq e^{-2\tanh(\frac{s}{2})H}
\end{equation}
for all $s>0$.
\end{Thm}
\proof
Consider an inclusion of intervals $I\Subset \tilde I$ with $\tilde I = \Iu$, $I$ symmetric with respect to the vertical axis and $\ell(\tilde I, I)=s$. By Prop. \ref{fin} we have $T_{\tilde I, I} =  e^{-sL_0}\D_{2}^{is/2\pi }$ thus
\begin{equation*}
T_{\tilde I, I}T^*_{\tilde I, I} =  e^{-2sL_0}\ .
\end{equation*}
On the other hand by Prop. \ref{fin2} we have
\begin{equation*}
T_{\tilde I, I}T^*_{\tilde I, I} = e^{-aH}e^{-2a'H'}e^{-aH}
\end{equation*}
where $a>0$ and $a'>0$ satisfy $\tau_{-a'}\tau_a\tilde I = I$. By equation \eqref{fa} we have $a' = \sinh(s)/2$ and $a = \tanh(s/2)$, so we have formula \eqref{m1}. 

Equation \eqref{m1} immediately entails $e^{-2sL_0} \leq e^{-2\tanh(\frac{s}{2})H}$. 
\endproof
Note that the inequality 
\begin{equation}\label{m2'}
e^{-sL_0} \leq e^{-\tanh(\frac{s}{2})H}
\end{equation}
follows from \eqref{m2} because the square root is an operator-monotone function.\footnote{The inequality \eqref{m2'} does not follow from $L_0\geq \frac12 H$ because the exponential is not operator monotone.}
Note also that the equation
\begin{equation*}
e^{-2sL_0} = e^{-\tanh(\frac{s}{2})H'}e^{-\sinh(s)H}e^{-\tanh(\frac{s}{2})H'}
\end{equation*}
follows by \eqref{m1} by applying a conjugation by a  $\pi$-rotation on both sides.
\smallskip

\noindent{\it Remark.} We may formally analytically continue the parameter $s$ in formula \eqref{m1} to the imaginary axis and get the equality
\begin{equation}\label{m1''}
e^{-2isL_0} = e^{i\tan(\frac{s}{2})H}e^{i\sin(s)H'}e^{i\tan(\frac{s}{2})H}\ ,
\end{equation}
in particular
\begin{equation}
e^{-i\pi L_0} = e^{iH}e^{iH'}e^{iH}\ .
\end{equation}
Indeed equation \eqref{m1''} holds true as one can check that it corresponds to an identity in the group $SL(2,\mathbb R)$. One can then use \eqref{m1''} to give an alternative derivation of the identity \eqref{m2}. This shows that equation \eqref{m2} holds for {\em all} unitary representations of $G$, without assuming positive energy, although the involved operators become unbounded in the general case.

\subsection{More general embeddings}

With $U$ a positive energy representation of $\Mob$ as above, we shall need to estimate the nuclear norm of the more general embedding operators
\[
T^U_{I,I_0}(\l)= T^U_{I,I_0}(\l) \equiv \Delta_I^\l \Delta_{I_0}^{-\l}, \quad 0<\l<1/2,
\]
associated with an inclusion of intervals $I_0\Subset I$. Clearly
\[
T^U_{I,I_0}=T^U_{I,I_0}(1/4).
\]
\begin{Prop}\label{l4}
For an inclusion of intervals $I_t\subset I$ with $I_t \equiv I_{t,t} = \tau'_{-t}\tau_t I$ as above, thus $\ell'(I,I_t) = t$, we have
\[
T_{I,I_t}(\l) = e^{-i\cos(2\pi\l)t H'_I}\big( e^{- \sin(2\pi \l)t H'_I}
e^{- \sin(2\pi \l)t H_I}\big) e^{i\cos(2\pi\l)t H_I}  e^{-it H_I}e^{it H'_I}\ .
\]
\end{Prop} 
\proof
\begin{equation}\label{fft2}
\begin{aligned}
T^U_{I,I_t}(\l) &\equiv \D_I^{\l}\D_{I_t}^{-\l} \\
&= \D_I^{\l}e^{-it H'_I}e^{it H_I}\D_I^{-\l}e^{-it H_I}e^{it H'_I}\\
&=\big(\D_I^{\l}e^{-it H'_I}\D_I^{-\l}\big)\big(\D_I^{\l}e^{it H_I}\D_I^{-\l}\big)e^{-it H_I}e^{it H'_I}\\
&=e^{-i(e^{-2\pi i\l})t H'_I} e^{i(e^{2\pi i\l})t H_I} e^{-it H_I}e^{it H'_I}\\
&=e^{-i(\cos(2\pi\l) - i \sin(2\pi\l))t H'_I}e^{i(\cos(2\pi\l) + i \sin(2\pi\l))t H_I} e^{-it H_I}e^{it H'_I}\\
&=e^{-i\cos(2\pi \l)t H'_I} e^{- \sin(2\pi \l)t H'_I}
e^{- \sin(2\pi \l)t H_I} e^{i\cos(2\pi \l)t H_I}  e^{-it H_I}e^{it H'_I}
\end{aligned}
\end{equation}
\endproof
\begin{Cor} 
$\| T_{I,I_t}(\l) \|_1 = \| T_{I,I_{\sin(2\pi \l)t}} \|_1\ .$
\end{Cor}
\proof
Immediate because by Proposition \ref{l4} the operator $T_{I,I_{\sin(2\pi \l)t}}$ is obtained by  left and  right multiplication of $T_{I,I_t}(\l)$ by unitary operators.

\section{A further operator inequality}

We shall need a version of an inequality proved in \cite{BDL2}, see Appendix \ref{bdl2}.

\begin{Prop} Let $U$ be a positive energy, unitary representation of $G$. We have
\begin{equation}\label{kdc}
||e^{-\tan(2\pi\l) d_I H}\Delta_I^{-\l}||\leq 1\ , \qquad 0<\l<1/4\ ,
\end{equation}
where $I$ is an interval of $\mathbb R$ with usual length $d_I$.
\end{Prop}
\proof
Assume first that $U$ is a representation of $\PSL$. We may consider $U$ as a representation of $\PSL\times\PSL$ which is trivial on the second component. As $\PSL\times\PSL$ is the symmetry group of a two-dimensional M\"obius covariant net on the 2-dimensional Minkowski spacetime, following the comments in Appendix \ref{bdl2} we get the bound \eqref{kdc}.

If $U$ is a representation of a finite $n$-cover of $\PSL$, we consider the $n$-time tensor product representation $U\otimes\cdots\otimes U$, which is a representation of $\PSL$. The the inequality \eqref{kdc} follows also in this case.

Finally, if $U$ is any representation of $G$, the result follows by a continuity argument by using Lemma \ref{GLW}.
\endproof
Note that the inequality \eqref{kdc} gives
\[
\Delta_I^{-\l}e^{-2\tan(2\pi\l) d_I H}\Delta_I^{-\l}\leq 1
\]
namely
\[
e^{-2\tan(2\pi\l) d_I H}\leq \Delta_I^{2\l}\ .
\]
In particular, if $I=I_2$ is the interval $(-1,1)$ of the real line, we have $e^{-4\tan(2\pi\l) H}\leq \Delta_2^{2\l}$. By conjugating both members of the inequality with the modular conjugation $J_{I_2}$ (ray inversion map) we get $e^{-4\tan(2\pi\l) H'}\leq \Delta_2^{-2\l}$, namely
\begin{equation}
e^{-4\tan(2\pi\l) H}\leq \Delta_2^{2\l}\leq e^{4\tan(2\pi\l) H'} 
\end{equation}
and, by rescaling with the dilation unitaries we obtain the inequality
\begin{equation}
e^{-2\tan(2\pi\l) d_I H}\leq \Delta_I^{2\l}\leq e^{2\tan(2\pi\l)\frac{1}{d_I} H'} 
\end{equation}
In particular, evaluating at $\l=1/8$, we get
\begin{equation}
e^{-2d_I H}\leq \Delta_I^{1/4}\leq e^{\frac{2}{d_I} H'} \  .
\end{equation}
\section{Modular nuclearity and $L^2$-nuclearity}

\subsection{Basic abstract setting}
\label{modnuc}

We now introduce the concept of $L^2$-nuclearity in abstract setting of inclusions of von Neumann algebras with a distinguished cyclic and separating vector. This will immediately provide the notion of $L^2$-nuclearity in conformal QFT by considering an inclusion of local observable von Neumann algebras and the vacuum vector.

Let $M$ be a von Neumann algebra  on a Hilbert space $\H$ and 
cyclic and separating unit vector $\Omega$. We set
\[
L^{\infty}(M)=M,\qquad L^{2}(M)=\H , \quad \quad L^{1}(M)= M_*\ .
\]
Then we have the embeddings
\[
\xymatrix{
L^{\infty}(M)  \ar[rr]^{x\to (x\Omega, J\,\cdot\, \Omega)}_{\Phi^M_{\infty,1}} \ar[ddr]^{\Phi^M_{\infty,2}}_{x\to \D^{1/4}x\Omega}& 
&  L^{1}(M)  \\
&\\
& L^{2}(M) \ar[uur]_{ \xi\to (\xi , J\,\cdot\,\Omega) }^{\Phi^M_{2,1}} }
\]
All embeddings are bounded with norm one.

Let now $N\subset M$ be an inclusion of von Neumann algebras with 
cyclic and separating unit vector $\Omega$.

Given $p,q=1,2,\infty$, $p\geq q$, we shall say that {\em $L^{p,q}$-nuclearity} 
holds for $N\subset M$, with respect to $\Omega$, if 
$\Phi^M_{p,q}|_{N}$ is a nuclear operator.
\footnote{Recall that a linear operator $A: X \to Y$ between Banach spaces $X,Y$ is called nuclear if 
there exist sequences of elements $f_k\in X^*$ and $y_k\in Y$ such that $\sum_k ||f_k||\, ||y_k||<\infty$ and $Ax=\sum_k f_k(x)y_k$. 
The infimum $||A||_1$ of $\sum_k ||f_k||\, ||y_k||$ over all possible choices of $\{f_k\}$ and $\{y_k\}$ as above is the nuclear  
norm of $A$.}

$L^{\infty,2}$ nuclearity has played an important r\^{o}le and has 
been named {\em modular nuclearity} \cite{BDL1}.

As $\Phi^M_{\infty,1}=\Phi^M_{2,1}\Phi^M_{\infty,2}$, 
we have 
\[
||\Phi^M_{\infty,1} |_N||_1 \leq ||\Phi^M_{2,1}||\cdot ||\Phi^M_{\infty,2}|_N||_1
\leq ||\Phi^M_{\infty,2}|_N||_1 \ ,
\]
where $||\cdot ||_1$ denotes the nuclear norm.  Thus
modular nuclearity implies $L^{\infty,1}$ nuclearity. Indeed the 
following result from \cite{BDL1,BDL2} holds:

\begin{Prop}\label{split1} {\rm \cite{BDL1,BDL2}}
Modular nuclearity implies $L^{\infty,1}$ nuclearity and 
$||\Phi^M_{\infty,1} |_N||_1 \leq ||\Phi^M_{\infty,2}|_N||_1$.
    
Conversely if $\Phi^M_{\infty,2}|_N$ is nuclear of type $s$, then 
also $\Phi^M_{\infty,1}|_N$ is nuclear of type $s$.
\end{Prop}
Of course the first part of the above statement is immediate by the above diagram: indeed $\Phi^M_{\infty,1} |_N =\Phi^N_{2,1}\cdot\Phi^M_{\infty,2}|_N$ and $|| \Phi^N_{2,1}||\leq 1$.
\begin{Prop}\label{split2} {\rm \cite{BDL1,BDL2}} If $N$ or $M$ is a factor and $\Phi^M_{\infty,1} |_N$ is nuclear (thus if modular nuclearity holds) then $N\subset M$ is a split inclusion.	
\end{Prop}
\proof We recall the short proof. By definition $\Phi^M_{\infty,1} |_N$ nuclear means that there exist sequences of elements $\varphi_k\in N^*$ and $\psi_k\in {M'}_* \, (\simeq L^1(M))$ such that $\sum_k ||\varphi_k||\ ||\psi_k||<\infty$ and 
\[
\omega(nm')=\sum_k \varphi_k(n)\psi_k(m')\ , \quad n\in N,\, m' \in M' \ .
\]
where $\omega\equiv (\,\cdot\, \Omega,\Omega)$. As $\Phi^M_{\infty,1} |_N$ is normal the $\varphi_k$ can be chosen normal (take the normal part). Thus the state $\omega$ on $N\odot M'$ extends to $N\otimes M'$ and this gives the split property.
\endproof
The above proposition also holds in the non-factor case
with $N$ and $M'$ generating algebraically a tensor product (which is 
automatic in the factor case) if $M$ is properly infinite.	
	
Consider now the commutative diagram
\[
\CD
L^{\infty}(N)    @> \phantom{xxxxx}\Phi^M_{\infty,1}|_N\phantom{xxxxx}
>>  L^{1}(M) 
\\ @V\Phi^N_{\infty,2} V  V 
@A  A\Phi^M_{2,1}  A  \\  
L^{2}(N)   @>\phantom{xx} T_{M,N}\equiv\D_M^{1/4}\D_N^{-1/4}\phantom{xx}>>L^{2}(M)
  \endCD
\]
Recall that that the operator $\D_M^{1/4}\D_N^{-1/4}$ is densely 
defined with norm one; its closure $T_{M,N}$ here above is the 
canonical embedding of $L^2(N)$ into $L^2(M)$.

We shall now consider the condition that $T_{M,N}$ be a nuclear 
operator that we call the {\em $L^2$-nuclearity condition}.

\begin{Prop}
$L^2$-nuclearity implies modular nuclearity and  $||\Phi^M_{\infty,2} |_N||_1 \leq ||T_{M,N}||_1$.
\end{Prop}
\proof
Immediate because $\Phi^M_{\infty,2}|_N =T_{M,N}\cdot\Phi^N_{\infty,2}$ and $||\Phi^N_{\infty,2}||\leq 1$.
\endproof
\smallskip

\noindent {\it Remark.} As shown in \cite{BDL1}, the split property for an inclusion of von Neumann algebras as in Section \ref{modnuc} implies modular compactness, i.e. $\Phi^M_{\infty,q}|_{N}$ is a compact operator, $q=1$ or $2$, (independently of the choice of the vector $\Omega$).

It is natural to wonder whether the split property implies the compactness of the operators $T_{M, N}$. We shall see in Section \ref{wedges} that this is not the case by computing the operator $T_{M, N}$ for inclusions of local von Neumann algebras associated to wedge regions on Minkowski spacetime.

\subsection{Varying the exponent}

We shall consider the condition
\[
\| T_{M,N}(\l) \|_1 <\infty 
\]
with $T_{M,N}(\l) \equiv \Delta_M^\l\Delta_N^{-\l}$ for general exponents $0<\l<1/2$.
Note that $\| T_{M,N}(\l) \| \leq 1$ by a standard interpolation argument \cite{BDL1}.

Consider the map $\Xi^M_\l : M\to\H$
\[
\Xi^M_\l : x\in M\mapsto \Delta_M^\l x\Omega \in\H
\]
thus $\Xi^M_{1/4} = \Phi^M_{\infty,2}$. We have $|| \Xi^M_\l || \leq 1$ if $0\leq \l\leq 1/2$. Since
\[
\Xi^M_{\l}|_N = T_{M,N}(\l)\cdot\Xi^N_{\l}
\]
we have
\[
|| \Xi^M_{\l}|_N ||_1 \leq || T_{M,N}(\l) ||_1\ .
\]

\section{Conformal nets and nuclearity conditions}
\label{cnets}

We now consider a {\em M\"obius covariant net} $\A$. Namely $\A$ is a map
\[
I\mapsto\A(I)
\]
from the set of proper intervals of $S^1$ to von Neumann algebras on a fixed Hilbert space $\H$. We assume: 

\item {\it Isotony}, i.e. $I\subset\tilde I\Rightarrow \A(I)\subset\A(\tilde I)$; \item{\it M\"obius covariance}, i.e. there exists a unitary representation $U$ of $\Mob$ on $\H$ such that $U(g)\A(I)U(g)^*=\A(gI)$; 
\item{\it Positive energy}, i.e. the conformal Hamiltonian $L_0$, the generator of the one-parameter rotation subgroup, is positive; 
\item{\it Vacuum vector}, i.e. there is a $U$-invariant unit vector $\Omega$; \item{\it Irreducibility}, i.e. $\bigvee_I\A(I)=B(\H)$ and $\bigcap_I\A(I)=\mathbb C$. We refer e.g. to \cite{BGL1,DLR} for what we need here. We do not assume locality nor diffeomorphism covariance.

It follows that $\Omega$ is cyclic and separating for each fixed von Neumann algebra $\A(I)$, so the modular operator $\Delta_I$ is defined. Moreover $\Delta_I^{it}=Z(t)U(\L_I(-2\pi t))$. Here $Z$ is a one-parameter group of internal symmetries ($Z(t)\A(I)Z(-t)=\A(I)$ and $Z(t)\Omega =\Omega$). If the net $\A$ is local, then $Z$ is trivial. Note that the operator $T_{\tilde I,I}(\l)\equiv\D^{\l}_{\tilde I}\D^{-\l}_{I}$ here below is equal to $e^{-8\l\pi K_{\tilde I}}e^{8\l\pi K_I}$ also in the non-local case as $Z$ does not depend on the interval and thus the corresponding factors cancel out.

Consider the following nuclearity conditions for $\A$.

\item {\em Trace class condition}: $\Tr(e^{-sL_0})<\infty$, $s>0$;

\item {\em $L^2$-nuclearity}: $||T_{\tilde I,I}(\l)||_1<\infty$, $\forall I\Subset \tilde I$, $0<\l<1/2$;

\item {\em Modular nuclearity}: $\Xi_{\tilde I,I}(\l):x\in\A(I)\to\D_{\tilde I}^{\l}x\Omega\in\H$ is nuclear $\forall I\Subset \tilde I$, $0<\l<1/2$;

\item {\em Buchholz-Wichmann nuclearity}: $\Phi_I^{\rm BW}(s):x\in\A(I)\to e^{-sH} x\Omega\in\H$ is nuclear, $I$ interval of $\mathbb R$, $s>0$ ($H$ the generator of translations);

\item {\em Conformal nuclearity}: $\Psi_I(s):x\in\A(I)\to e^{-sL_0} x\Omega\in\H$ is nuclear, $I$ interval of $S^1$, $s>0$.

We shall show the following chain of implications:
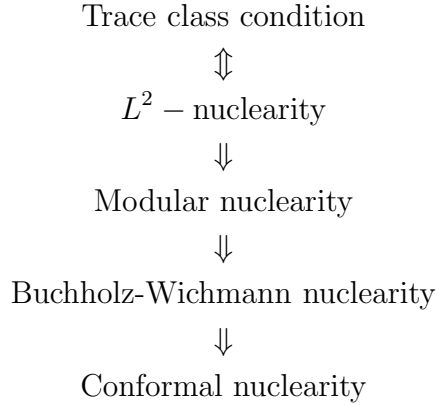
\begin{gather*}
\text{\rm Trace class condition}\\
\Updownarrow \\
L^2 -\text{\rm nuclearity}\\
\Downarrow\\
\text{\rm Modular nuclearity}\\
\Downarrow \\
\text{\rm Buchholz-Wichmann nuclearity}\\
\Downarrow \\
\text{\rm Conformal nuclearity}
\end{gather*}
Where all the conditions can be understood for a specific value of the parameter, that will be determined, or for all values in the parameter range.
\smallskip

\noindent {\it Remark.} The implication ``BW-nuclearity $\Rightarrow$ Modular nuclearity" also holds true if one assumes ``nuclearity of higher order", e.g. type $s$ \cite{BDL2}. The argument goes through the following chain of implications: ``BW-nuclearity of type $s$ $\Rightarrow$ $\Phi^{\A(\tilde I)}_{\infty,1}|_{\A(I)}$ is nuclear of type $s$ $\Rightarrow$ $\Phi^{\A(\tilde I)}_{\infty,2}|_{\A(I)}$ is nuclear of type $s$, with $I\Subset \tilde I\,$''.
\smallskip

We have already discussed the implications ``Trace class condition $\Leftrightarrow L^2$-nuclearity $\Rightarrow$ Modular nuclearity".

\subsection{Modular nuclearity $\Rightarrow$ BW-nuclearity}

Equation \eqref{kdc} gives
$||e^{-\tan(2\pi\l) d_I H}\Delta_I^{-\l}||\leq 1 $ for all $0<\l<1/4$, so the following holds:
\begin{Prop}\label{bwm}
Let $I_0\Subset I$ be a an inclusion of intervals of $\mathbb R$. We have
\[
||\Phi_{I_0}^{\rm BW}\big(\tan(2\pi\l) d_I\big)||_1 \leq ||\Xi_{I,I_0}(\l)||_1
\] where $d_I$ is the length of $I$, $0<\l<1/4$.
\end{Prop}
\proof
With $X\in\A(I_0)$ we have
\[
\Phi_{I_0}^{\rm BW}(s)X\Omega = e^{-sH}X\Omega
= \big(e^{-sH}\Delta_{I}^{-\l}\big)\Delta_{I}^\l X\Omega 
\]
thus 
\[
\Phi_{I_0}^{\rm BW}\big(\tan(2\pi\l)d_I\big) = \big(e^{-\tan(2\pi\l) d_I H}\Delta_{I}^{-\l}\big)\cdot \Xi_{I,I_0}(\l)
\]
and so
$
||\Phi_{I_0}^{\rm BW}\big(\tan(2\pi\l) d_I \big)||_1 \leq ||\Xi_{I,I_0}(\l)||_1
$
as desired
\endproof
\subsection{Quantitative and asymptotic estimates}

At this point we have the following chain of inequalities: 
\begin{align*}
||\Phi_{I_0}^{\rm BW}\big(\tan(2\pi\l) d_I\big)||_1 & \leq ||\Xi_{I,I_0}(\l)||_1  \\
& \leq ||T_{I,I_0}(\l)||_1 &\\
& = ||T_{I,I_1}||_1  \quad\qquad\ell'(I,I_1) = \sin(2\pi\l)\ell'(I,I_0)
\\
& =  \Tr(e^{-s L_0}), \qquad s = \ell(I,I_1)
\end{align*}
Note that $s = 2\sinh^{-1}\!\!\big(\ell'(I,I_1)\big)=2\sinh^{-1}\!\!\big(\sin(2\pi\l)\ell'(I,I_0)\big)$ by Prop. \ref{ll'}.

As $\l\to 0^+$ one has $\tan(2\pi\l) \sim 2\pi\l$ and $s\sim 4\pi \l \ell'(I,I_0)$ so we have the asymptotic inequality
\begin{equation}\label{estimate}
||\Phi_{I_0}^{\rm BW}(a)||_1 \leq \Tr(e^{- (2\ell'(I,I_0)/d_I)a L_0}), \quad a\to 0^+\ .
\end{equation}

Here below we have our last estimate that will give a relation to conformal nuclearity.

\subsection{BW-nuclearity $\Rightarrow$  Conformal nuclearity}

By equation \eqref{m2} there exists a bounded operator $B$ with norm $||B||\leq 1$  such that
\begin{equation*}
e^{-sL_0} = Be^{-\tanh(\frac{s}{2})H}
\end{equation*}
therefore
\begin{equation}\label{cbw}
\Psi_I(s) = B\Phi_I^{\rm BW}(\tanh(s/2))
\end{equation}
and so we have
\begin{Prop} $ ||\Psi_I(s)||_1 \leq ||\Phi_I^{\rm BW}(\tanh(s/2))||_1$ .
\end{Prop}
One more consequence of eq. \eqref{cbw} is the following.
\begin{Cor} If the split property holds for $\A$ then ``conformal compactness" holds, namely $\Psi_I(s)$ is a compact operator for all $s>0$ and all intervals of $S^1$.
\end{Cor}
\proof
The split property implies ``modular compactness" ($\Xi_{\tilde I,I}(\l)$ is compact)  \cite{BDL1} and ``modular compactness" implies ``BW-compactness"($\Phi_I^{\rm BW}(s)$ is compact) by eq. \eqref{kd} \cite{BDL2}. The corollary thus follows by equation \eqref{cbw}.
\endproof

\subsection{Deriving the split property}
The implication ``trace class condition $\Rightarrow$ modular nuclearity'' has the following corollary (cf. \cite{DLR}), namely a simple and direct derivation of the split property.
\begin{Cor}
Let $\A$ satisfy the trace class condition at a fixed $s_0 >0$, i.e. $\Tr(e^{-s_0 L_0})<\infty$. Then the distal split property holds, more precisely $\A(I)\subset\A(\tilde I)$ is a split inclusion if $\tilde I\supset I$ and $\ell(\tilde I, I)> s_0$.
\end{Cor}
\proof
By eq. \eqref{TT} we have $L^2$-nuclearity holds if $\ell(\tilde I, I) >s_0$. Thus modular nuclearity holds for the inclusion $I\subset\tilde I$ by Prop. \ref{split1}. The split property then holds for $\A(I)\subset\A(\tilde I)$ by Prop. \ref{split2}.
\endproof

\section{Constructing KMS states}

Let $\A_0$ be an irreducible net of von Neumann algebras on $\mathbb R$, which is translation covariant with positive energy and vacuum vector (with the Reeh-Schlieder cyclic and separating properties). We denote by the same symbol $\A_0$ the quasi-local C$^*$-algebra. i.e. the norm closure of $\cup_I \A(I)$ as $I$ varies in the bounded intervals of $\mathbb R$. Let $\mathfrak A\subset \A_0$ the C$^*$-algebras of elements  with norm continuous orbit, namely
\[
\mathfrak A = \{X\in\A_0: \lim_{t\to 0}||\tau_t(X)-X||=0\}
\]
where as above $\tau$ is the translation automorphism one-parameter group.

Consider the BW-nuclearity condition for $\A_0$ with the asymptotic bound 
\begin{equation}\label{bound}
||\Phi^{BW}_I(\beta)||_1 \leq e^{cr^m\beta^{-n}}, \quad \b\to 0^+ \ ,
\end{equation}
where $c$, $m$ and $n$ are positive constant and $r$ is the length of $I$.

Let now $\A$ be a M\"obius covariant net on $S^1$ and denote by $\A_0$ its restriction to $\mathbb R \simeq S^1\setminus\{-1\}$.
\begin{Thm}\label{BJT}
{\rm \cite[Thm. 3.4]{BJ}.}  
If BW-nuclearity 
holds for $\A$ with the asymptotic bound \eqref{bound}, then  
for every $\beta >0$ there exists state $\varphi_{\beta}$ on $\mathfrak A$ which is $\beta$-KMS with respect to $\tau$.
\end{Thm}

With $\A$ a conformal net as above, fix an interval $I_0$ of the real line. Let $I\equiv \l^{-1} I_0$, where $\l>0$. We now show that, because of dilation covariance, the behavior of the BW nuclearity index as $I$ increases at fixed inverse temperature $\b$ is the same of as $I$ is fixed and $\b\to 0$.

\begin{Lemma}
$||\Phi^{BW}_I(s)||_1 =  ||\Phi^{BW}_{I_0}(\l s)||_1$.
\end{Lemma}
\proof
We have 
\begin{multline*}
e^{-s\l H}\A(I_0)\Omega =e^{-s\l H}\A(\l I)\Omega
= e^{-s\l H}D(\l)\A(I)\Omega\\
= D(\l)\big(D(\l)^*e^{-s\l H}D(\l)\big)\A(I)\Omega
= D(\l) e^{-s H}\A(I)\Omega
\end{multline*}
where $D(e^t)=U(\L_{\mathbb R^+}(t))$, and this readily implies the statement.
\endproof

\begin{Thm} Let $\A$ be a M\"obius covariant net on $S^1$. If the trace class condition holds for $\A$ with the asymptotic bound
\[
\Tr(e^{-sL_0})\leq e^{{\rm const.}\frac{1}{s^{\alpha}}}, \quad s\to 0^+
\]
for some $\alpha>0$, then the bound \eqref{bound} holds with $m=n=\alpha$.

As a consequence for every $\b>0$ there exists a translation $\b$-KMS states on $\mathfrak A$.
\end{Thm}
\proof
Immediate by the Buchholz-Junglas Theorem \ref{BJT}, the estimate \eqref{estimate} and the above Lemma.
\endproof
\noindent
In case the net $\A$ is diffeomorphism covariant, a direct construction of translation KMS states has been pointed out to us by Mih\'aly Weiner (work in progress).

\section{A look at nets in higher dimension}
\label{Mink}

Let $\A$ be a net of von Neumann algebras on the Minkowski spacetime. We shall need only isotony, positive energy and vacuum vector with Reeh-Schlieder property for wedge regions (neither locality or Bisognano-Wichmann property will be used).

We shall say that $L^2$-nuclearity holds for $\A$ if the inclusion of von Neumann algebras $\A(\O)\subset\A(\tilde\O)$ satisfies $L^2$-nuclearity with respect to the vacuum vector $\Omega$ for double cones $\O\Subset\tilde\O$, namely
\[
||T_{\tilde\O,\O}||_1<\infty
\]
where $T_{\tilde\O,\O}\equiv \D_{\tilde\O}^{1/4}\D_{\O}^{-1/4}$ and $\D_{\O}$ is the modular operator associated with $(\A(\O),\Omega)$. We shall consider the operator $T_{\tilde\O,\O}$ also in the case of different regions (wedges).

\subsection{On wedge inclusions on Minkowski spacetime}
\label{wedges}

Let then $\A$ be a net on the Minkowski spacetime as above.
We consider the wedge $W\equiv \{x: x_1 >|x_0|\}$ and a sub-wedge $W_a\equiv \{x: x_1 >|x_0| + a \}$ with $a>0$. Analogously as in formula \eqref{fft} have:

\begin{Prop} $T_{W, W_a} = e^{-aP_0}e^{-ia P_1}$,
where $P_0$, $P_1$ are the generator of time translations (energy operator) and space translation in the $x_1$-direction (momentum operator).
\end{Prop}
\proof
Clearly $\A(W_a)= e^{iaP_1}\A(W)e^{-iaP_1}$. Therefore
\[
\D_{W_{a}}^{-1/4} = e^{iaP_1}\D_{W}^{-1/4}e^{-iaP_1}=e^{i\frac{a}{2}(P_0 + P_1)}e^{-i\frac{a}{2}(P_0 - P_1)}\D_{W}^{-1/4}e^{i\frac{a}{2}(P_0 - P_1)}e^{-i\frac{a}{2}(P_0 + P_1)}
\]
Since the light-like translations in the $x_0 \pm x_1$ direction maps  $W$ into itself for positive/negative translations and the corresponding unitary groups have positive generators $P_0 \pm P_1$, we can proceed similarly as in \eqref{fft}:
\begin{align*}
\D_W^{1/4}\D_{W_a}^{-1/4} =& \D_W^{1/4}e^{i\frac{a}{2} (P_0 + P_1)}e^{-i\frac{a}{2}(P_0 - P_1)}\D_{W}^{-1/4}e^{i\frac{a}{2}(P_0 - P_1)}e^{-i\frac{a}{2}(P_0 + P_1)}\\
= & \big(\D_W^{1/4}e^{i\frac{a}{2} (P_0 + P_1)}\D_W^{-1/4}\big)\big(\D_W^{1/4}e^{-i\frac{a}{2}(P_0 - P_1)}\D_{W}^{-1/4}\big)e^{i\frac{a}{2}(P_0 - P_1)}e^{-i\frac{a}{2}(P_0 + P_1)}\\
= & e^{-\frac{a}{2} (P_0 + P_1)}e^{-\frac{a}{2}(P_0 - P_1)}e^{i\frac{a}{2}(P_0 - P_1)}e^{-i\frac{a}{2}(P_0 + P_1)}\\
= & e^{- a P_0}e^{-ia P_1}
\end{align*}
where we have used Borchers commutation relations \cite{B}.
\endproof
\smallskip

\noindent {\it Remark.} Suppose now $\A$ to be the net generated by a free scalar field of mass $m>0$ on the two dimensional Minkowski spacetime. It is known that in this case the inclusion $\A(W_a)\subset \A(W)$ is split if $a>0$, see \cite{DF}. As the spectrum of $P_0$ contains a continuos part, $e^{-aP_0}$ is definitely not compact.

Assuming the split property, the local von Neumann algebras are isomorphic to the unique Connes-Haagerup injective factor of type $III_1$. Moreover all split inclusion $N\subset M$ with $N$ and $M$ the injective $III_1$-factor are isomorphic. Therefore we have shown that there exists joint cyclic and separating vectors $\xi_1$ and $\xi_2$ for this inclusion such that $L^2$-nuclearity holds w.r.t. $\xi_1$ but even $L^2$-compactness fails to hold w.r.t. $\xi_2$.

Note also that, if we consider the associated one-particle real Hilbert space structure, that is a unitary, massive representation of the two-dimensional Poincar\'e group, then as above $T_{W, W_a}$ is not compact; moreover, as
\[
||T_{W, W_a}||= ||e^{-aP_0}|| = e^{-am}<1
\]
we have examples where the uniform norm $||T_{W, W_a}||$ is arbitrarily small and the corresponding real Hilbert space inclusion is not split.

\subsection{$L^2$-nuclearity for the scalar, massless, free field}

With $\O$ a double cone in the Minkowski spacetime $\mathbb R^{d+1}$, we denote here by $\A(\O)$ the local von Neumann algebra associated with $\O$ by the $d+1$-dimensional  scalar, massless, free field.

With $I$ an interval of the time-axis $\{x=\langle x_0,{\mathbf x}\rangle: {\mathbf x} = 0\}$ we set
\[
\A_0(I)\equiv \A(\O_I)
\]
where $\O_I$ is the double cone $I''\subset\mathbb R^{d+1}$, the causal envelope of $I$. Then $\A_0$ is a translation-dilation covariant net on $\mathbb R$. $\A_0$ is local if $d$ is odd and twisted local if $d$ is even. Moreover $\A_0$ extends to a M\"obius covariant 
net on $S^1$ ($d$ odd) or on the double cover of $S^1$ ($d$ even). For simplicity here below we treat the case of $d$ odd; the case $d$ even may be dealt analogously.

The one particle Hilbert space $\K$ is (complex span of) the completion of $S(\mathbb R^{d+1})$ equipped with the scalar product
\[
(f,g) = \frac{1}{(2\pi)^d}\int \frac{{\rm d}^d{\mathbf p}}{2|{\mathbf p}|}\tilde{\bar f}(-|{\mathbf p}|,-{\mathbf p})\tilde g (|{\mathbf p}|,{\mathbf p})\ .
\]
Consider a distribution $H$ on $\mathbb R^{d+1}$ of the form
\begin{equation}\label{distr}
H(x) = \frac{\partial^{k_1}}{\partial x_1^{k_1}}\dots
\frac{\partial^{k_d}}{\partial x_d^{k_d}}\delta(\mathbf x)h(x_0)
\end{equation}
where $\delta(\mathbf x)$ is the Dirac function concentrated on the on the time-axis and $h$ is a real smooth function in $S(\mathbb R)$. Call $k\equiv k_1 + \cdots k_d$ the order of $H$.

The anti-Fourier transform $\tilde H$ is a homogeneous degree $k$ polynomial in $p_1,\dots p_d$ times $\tilde h(p_0)$.
Denote by $\T_k$ the liner span of distributions as in \eqref{distr} with order $k$. So  $\T_k\subset \K$ and the linear span $\T$ of the $\T_k$'s is dense in $\K$
(see \cite{HL}).
\begin{Lemma}
Given an interval $I\subset\mathbb R$, $\A(\O_I)$ is generated by the Weyl unitaries $W(H)$, as $H$ varies in $\bigcup_k\T_k$ with ${\rm supp}(H)\subset I$ (i.e. ${\rm supp}(h)\subset I$). 
\end{Lemma}
\proof
Denote by $\B(I)$ the von Neumann algebra generated by the Weyl unitaries $W(H)$, as $H$ varies in $\bigcup_k\T_k$ with ${\rm supp}(H)\subset I$. Clearly $\B$ is a translation-dilation covariant net on $\mathbb R$ with positive energy and $\B(I)\subset\A_0(I)$. 
By the above density of $\T$ in $\K$, the net $\B$ is cyclic on the vacuum vector of $\A$. By the Reeh-Schlieder theorem $\B(I)$ is cyclic on the vacuum. By the geometric expression of the vacuum modular group of $\A(\O_I)$ \cite{HL}, $\B(I)$ is globally invariant under the modular group of $\A(\O_I)$. By the Tomita-Takesaki theory $\B(I)=\A(\O_I)$.
\endproof
\begin{Lemma}\label{mult}
If $d$ is odd, then
\[
\A_0 =\bigotimes_{k=0}^{\infty} N_d(k)\A^{(k)}
\]
where $\A^{(k)}$ is the M\"obius covariant net on $S^1$ associated with the $k^{\rm th}$-derivative of the $U(1)$-current algebra and $N_d(k)$ is a multiplicity factor (see below).
\end{Lemma}
\proof
Let $U$ be the irreducible, positive energy representation of the Poincar\'e group with mass $m=0$ and zero helicity. Then $\A$ is associated with the second quantization of $U$ as in \cite{BGL3}. 

Then $U$ extends to an irreducible unitary representation of the $d+1$-dimensional conformal group $C_d\equiv SO_0(2,d+1)$ on the same Hilbert space, that we still denote by $U$. The subgroup of $C_d$ generated by time-translations, dilations and ray inversion is isomorphic to the M\"obius group ($=PSL(2,\mathbb R)$). Denote by $U_0$ the restriction of $U$ to this copy of $PSL(2,\mathbb R)$. Then $\A_0$ is associated to the second quantization of $U_0$. Indeed the real Hilbert subspace (in the one-particle Hilbert space) associated with $\O_I$ by the representation $U$ of $C_d$ is defined only in term of $U_0$ ($K_I$ generates a one-parameter unitary subgroup of $U_0$).

Clearly
\[
U_0 =\bigoplus_{k=1}^{\infty} N_d(k) U^{(k)}
\]
where $ U^{(k)}$ is the positive energy irreducible representation of $PSL(2,\mathbb R)$ with lowest weight $k$. As the second quantization net associated with $U^{(k)}$ is $\A^{(k)}$, we get the thesis.
\endproof
We now determine the multiplicity factor $N_d(k)$ in Lemma \ref{mult}. As in the case $d=1$ we have $\A_0 =\A^{(0)}\otimes\A^{(0)}$, cf. \cite{HL}, we may assume $d>1$.

Now if $H_1, H_2\in\T_k$ then
\begin{equation}\label{H}
(H_1,H_2)=  \frac{1}{(2\pi)^d}\int_0^{\infty} p_0^{2k + d-2}\tilde{\bar h}_1(-p_0)\tilde h_2(p_0){\rm d}p_0
\int_{S^{d-1}}\bar P_1({\mathbf p})P_2({\mathbf p})|_{|{\mathbf p}|=1} {\rm d}\sigma_{d-1}
\end{equation}
where $P_1$ and $P_2$ are the homogeneous degree $k$ polynomials in ${\mathbf p}\equiv\langle p_1,\dots p_d\rangle$ appearing in the Fourier anti-transform of $H_1$ and $H_2$ and $\sigma_{d-1}$ is the volume element of the unit $d-1$-dimensional sphere in $\mathbb R^d$.

\medskip

\noindent {\it Remark.} (Spherical harmonics, cf. \cite{K})
Let $m_d(k)$ be the number of monomials $p_1^{k_1}\dots p_d^{k_d} $ with order $k= k_1 +\dots + k_d$. Note that we have
\begin{equation}\label{md}
m_d(k) = \sum^k_{h=0} m_{d-1}(h)
\end{equation}
so $m_1(k)= 1$, $m_2(k)= k+1$, $m_3(k)= (k+1)(k+2)/2$, \dots, and we have
\[
m_d(k) \sim \frac{1}{(d-1)!}k^{d-1}, \qquad k\to\infty\ .
\]
Denote by $\P_k$ the functions on $S^{d-1}$ that are restrictions of homogeneous degree $k$ polynomials in $p_1,\dots p_d$. Clearly $\P_k\subset L^2(S^{d-1},{\rm d}\sigma_{d-1})$. Moreover $\P_{k-2}\subset\P_k$. The natural unitary representation of $SO(d)$ on $L^2(S^{d-1},{\rm d}\sigma_{d-1})$ leaves $\P_k$ globally invariant and the corresponding decomposition into irreducible subspace is
\[
L^2(S^{d-1},{\rm d}\sigma_{d-1})=\bigoplus_{k=0}^{\infty}(\P_k\ominus\P_{k-2})
\]
It follows that the irreducible $SO(d)$-subspaces of $L^2(S^{d-1},{\rm d}\sigma_{d-1})$ have dimension
\begin{equation}\label{pol}
{\rm dim}(\P_k\ominus\P_{k-2}) = m_d(k) - m_d(k-2)
= m_{d-1}(k-1) + m_{d-1}(k)\ .
\end{equation}
The elements of $\P_k\ominus\P_{k-2}$ are the harmonic spherical functions of degree $k$, namely a polynomial $P\in\P_k$ is orthogonal to $\P_{k-2}$ iff $P$, as function on $\mathbb R^d$, is annihilated by the Laplace operator.
\medskip

\noindent
Now $\T_k$ is an invariant subspace for the representation $U$ of $C_d$. The restriction $U_0$ of $U$ to $PSL(2,\mathbb R)$ obviously commutes with the restriction of $U$ to $SO(d)$. Therefore
the restriction of $U$ to $\T_k \ominus \T_{k-2}$ is the tensor product of an irreducible representation of $SO(d)$ and an irreducible, positive energy representation $V_k$ of $PSL(2,\mathbb R)$. We now show that the lowest weight of $V_k$ is equal to $k+(d-1)/2$.

Indeed by eq. \eqref{H} the Hilbert space $\K_k$ of $V_k$ is the completion of $S(\mathbb R)$ equipped with scalar product
\begin{multline*}
(h_1,h_2)_k \equiv  \int_0^{\infty} p_0^{2k + d-2}\tilde{\bar h}_1(-p_0)\tilde h_2(p_0){\rm d}p_0 \\
=\int_0^{\infty} p_0\tilde{\bar h}^{(k + \frac{d-3}{2})}_1(-p_0)\tilde h^{(k + \frac{d-3}{2})}_2(p_0){\rm d}p_0 =(h^{(k + \frac{d-3}{2})}_1,h^{(k + \frac{d-3}{2})}_2)_1
\end{multline*}
where $h^{(k)}$ is the $k$-derivative of $h$. Translation and dilation unitaries have a natural expression on $\K_k$. Clearly $\K_1$ is the one particle Hilbert space of the $U(1)$-current algebra. In the $x_0$ configuration space the ray inversion unitary corresponds to the geometric action $x_0\mapsto -1/x_0$ ($h(x_0)$ goes to $h(-1/x_0)$ multiplied by some power of $x_0$ depending on $k$). This suffices to conclude that $V_k$ the irreducible unitary representation of $SL(2,\mathbb R)$ with lowest weight $k+ (d-1)/2$.

By formula \eqref{pol} we thus have $N_d(k) = 0$ if $k<(d-1)/2$ and, as $k\to\infty$,
\[
N_d(\text{\small{$k\! +\! \frac{d\! -\! 1}{2}$}}) = {\rm dim}(\P_k\ominus\P_{k-2}) = m_{d-1}(k-1) + m_{d-1}(k)
\sim \text{\small{$\frac{2}{(d-2)!}$}}k^{d-2} \ .
\]
\begin{Lemma} We have
\[
\log\Tr(e^{-sL_0}) \sim \frac{2}{s^{d}}\qquad s\to 0^+ \ ,
\]
where $L_0$ is the conformal Hamiltonian of $\A_0$
\end{Lemma}
\proof
With $L^{(k)}$ the generator of rotations in the representation $U^{(k)}$ we have $\Tr(e^{-sL^{(k)}_0}) = \frac{e^{-sk}}{1- e^{-s}}$, therefore as $s\to 0^+$ we have:
\begin{multline*}
\Tr(e^{-sL^{U}_0}) = \frac{1}{1- e^{-s}}\sum_{k=(d-1)/2}^{\infty}N_d(k) e^{-sk}
\sim \frac{2}{(d-2)!}\frac{e^{-\frac{d-1}{2}s}}{(1- e^{-s})}\sum_{k=0}^{\infty}k^{d-2} e^{-sk}\\
\sim \frac{2}{(d-2)!}\frac{1}{(1- e^{-s})}\int_{0}^{\infty}t^{d-2} e^{-st}{\rm d}t
= \frac{2}{(d-2)!(1- e^{-s})}\frac{1}{s^{d-1}}\int_{0}^{\infty}t^{d-2} e^{-t}{\rm d}t\\
= \frac{2}{(1- e^{-s})}\frac{1}{s^{d-1}}\frac{\Gamma(d-1)}{(d-2)!}
= \frac{2}{(1- e^{-s})}\frac{1}{s^{d-1}}
\sim\frac{2}{s^{d}}
\end{multline*}
where $\Gamma$ is the Euler Gamma-function.

As $\log\Tr(e^{-sL_0}) \sim \Tr(e^{-sL^{U}_0})$ as $ s\to 0^+$ (see \cite[Appendix]{KL})
we have completed our proof.
\endproof
\noindent {\it Remark.} We give the explicit expression for the trace in the above estimate in the case $d=3$. In this case $N_3(k)=2k-1$, therefore
\begin{multline*}
\Tr(e^{-sL^{U}_0}) 
= \frac{1}{1- e^{-s}}\sum_{k=1}^{\infty}(2k-1) e^{-sk}
=  \frac{2}{1- e^{-s}}\sum_{k=1}^{\infty}ke^{-sk}
-\frac{1}{1- e^{-s}}\sum_{k=1}^{\infty}e^{-sk}\\
=  -\frac{2}{1- e^{-s}}\frac{{\rm d}}{{\rm d}s}\!\left(\frac{1}{1- e^{-s}}\right)
- \frac{e^{-s}}{(1- e^{-s})^2}
=  \frac{e^{-s}(1+ e^{-s})}{(1- e^{-s})^3}
= \frac{\cosh(s/2)}{4\sinh^3(s/2)}
\end{multline*}
that goes as $2s^{-3}$ when $s\to 0^+$.

\begin{Cor}
Let $\A$ be the net of von Neumann algebras on $\mathbb R^{d+1}$ associated with the free massless scalar field. Then $L^2$-nuclearity holds for $\A$, indeed
\[
||T_{\O_r,\O_1}||_1 \sim e^{{\rm const.}\frac{1}{(\log r)^{d}}}
\]
as $r\to 1^+$, where $\O_r$ is a double cone in $\mathbb R^d$, centered at the origin, of radius $r>1$.
\end{Cor}
By our general results (see Sect. \ref{cnets}) we find in particular the nuclearity estimates of Buchholz-Jacobi \cite{BJ}.

Note that we have proved the following.
\begin{Prop} Let $U$ be the unitary irreducible representation of the conformal group $SO(2,d+1)$ ($d$ odd) whose restriction to the $d+1$-dimensional Poincar\'e group is the positive energy, massless, zero helicity representation. Then the irreducible decomposition of the restriction of $U$ to the $PSL(2,\mathbb R)$ subgroup of $SO(2,d+1)$ generated by time-translation, dilations and ray inversion is given by
\[
U|_{PSL(2,\mathbb R)} = \sum_{k=0}^{\infty}
\big(m_{d-1}(k-1) + m_{d-1}(k)\big)U^{(k+1)}
\]
where $U^{(k)}$ is the $k$-lowest weight representation of $PSL(2,\mathbb R)$ and $m_d(k)$ is given by \eqref{pol}.
\end{Prop}

\section{Appendix}

\subsection{Inner distance}
\label{app}

Our estimates are based on the introduction of a certain inner distance for inclusions of intervals of $S^1$. Such concepts are discussed and clarified here below.

With $I \Subset\tilde  I$ the inner distance $\ell(\tilde I,
I)$ introduced in Section \ref{T1} can be more intrinsically defined as follows.  Let $w_1,w_2$ be the boundary points of $\tilde I$ and $z_1,z_2$ be the boundary points of $I$ in the
counterclockwise order, thus $w_1 \prec z_1\prec z_2\prec w_2$ (in the
counterclockwise order).  let $z$ be a point between $z_1$ and $z_2$. The reflection $r_{\tilde I}$ associated with $\tilde I$ maps $z$ in a point $z'\in \tilde I'$ and let $I_0$ 
be the interval $(z', z)$.
Choose $t$ such that $z'_1\equiv \L_{I_0}(t)z_1$ and $z'_2\equiv \L_{I_0}(t)z_2$ are conjugate under the reflection $r_{I_0}$.  Then $\ell(\tilde I,I)$ is the unique $s>0$ such that $\L_{I_0}(s)w_1 = z_1$.

It is easily seen that the inner distance satisfies the following properties:
\begin{itemize}   
    \item {\em Positivity}: $\ell(\tilde I,I)>0$ if $I \Subset\tilde  I$ 
    and all positive values are attained.
    
    \item {\em Monotonicity}: If $I_1 \Subset I_2\Subset
    I_3$ we have $\ell(I_3,I_1)>\ell(I_3,I_2)$ and
    $\ell(I_3,I_1)>\ell(I_2,I_1)$.
    
    \item {\em M\"obius invariance}: $\ell(\tilde I,I)=\ell(g\tilde I,gI)$
    for all $g\in G$.
    
    \item {\em Super-additivity}: $\ell(I_3,I_1)\geq\ell(I_3,I_2)+\ell(I_2,I_1)$
    if $I_1\Subset I_2\Subset I_3$. (See below).
\end{itemize}   
    
In section \ref{T3} we have considered a second inner distance $\ell'$ defined as follows. First $\ell'(\tilde I, I) = t$ if in the real line picture $\tilde I=\mathbb R_+$ and $I_t ={\tau'}_{-t}\tau_t \tilde I$, 
$(t>0)$.

Here $\tau_t$ is the translation by $t$ and $\tau'_{-t}$ is the conjugate of $\tau_t$ by the ray inversion
\[
\tau'_{-t} : x\mapsto \frac{x}{1 + tx}\ ,
\]
More generally we put for any $a, a' >0$
\[
\ell'(\tilde I, I) = \sqrt{aa'} \quad \text{if}\quad I =\tau'_{-a'}\tau_{a}\tilde I \ .
\]
Since conjugating $\tau_t$ and $\tau'_t$ by a dilation by $\l$ gives $\tau_{\l^{-1}t}$ and $\tau'_{\l t}$, the above is a M\"obius invariant definition for $I\subset\tilde I$ when $\tilde I$ is the positive half-line. 

In general, with $I_1\Subset I_2$ intervals of $S^1$, let $g\in\Mob$ be such that $g\tilde I = \Iu\, (\simeq \mathbb R_+)$; then we put
\[
\ell'(\tilde I, I)\equiv \ell'(g\tilde I, gI)\ .
\]
By the above comments, $\ell'$ is well-defined. It easily seen that also $\ell'$ is positive, monotone and M\"obius covariant. The two distances are related as follows.

\begin{Prop}\label{ll'}
$\ell' = \sinh(\ell/2)$. 
\end{Prop}
\proof
Given $t>0$ and an inclusion of intervals $I\subset \tilde I$ with $\ell'(\tilde I, I)= t$ we want to calculate $s\equiv \ell(\tilde I, I)$. We may assume that $\tilde I =\mathbb R_+$ and $I = I_{\l^{-1}t,\l t} \equiv\tau'_{\l^{-1}t}\tau_{\l t}\tilde I$. Then
\[
I =\tau'_{-\l^{-1}t}\tau_{\l t}\tilde I = \tau'_{-\l^{-1}t}(\l t, \infty) =\left(\frac{\l^{-1}t}{1+t^2} , \frac{1}{\l t}\right) \ .
\]
We may further choose $\l$ so that $I$ symmetric under ray inversion, namely $\l^{-1}=\sqrt{1 +t^2}$. thus
\begin{equation}\label{I}
I= \left(\frac{t}{\sqrt{1 +t^2}}, \frac{\sqrt{1 +t^2}}{t}\right)\ .
\end{equation}
Now in the real line picture the right semicircle $\Ir$ corresponds to the interval $(-1, 1)$, thus $s$ is determined by
\begin{equation}\label{mg}
\L_{(-1,1)}(s) 0 = \frac{t}{\sqrt{1 +t^2}} \ .
\end{equation}
Now 
\[
\L_{(-1,1)}(s) : x \mapsto \frac{x +1 - e^{-s}(x - 1)}{x +1 + e^{-s}(x - 1)}
\]
(see \cite{HL}) thus eq. \eqref{mg} gives 
\[
\frac{t}{\sqrt{1 +t^2}} = \frac{1 -  e^{-s}}{1 + e^{-s}} = \tanh(s/2) =\frac{\sinh(s/2)}{\sqrt{1 + \sinh^2(s/2)}}
\]
and this implies $t = \sinh(s/2)$.
\endproof

Note that $\L_{(-1,1)}(s)(0,\infty) =  I_{\l^{-1}t,\l t}
\equiv \tau'_{-\l^{-1}t}\tau_{\l t}$ 
with $\l = 1/\sqrt{1 + t^2}$, namely
\begin{equation}
\L_{(-1,1)}(s)(0,\infty)=
\tau'_{-t\sqrt{1 + t^2}}\tau_{t /\sqrt{1 + t^2}}(0,\infty)\ .
\end{equation}
In term of the inner distance $s$ the interval $I$ in \eqref{I} is given by
\begin{multline}\label{fa}
I= \big(\tanh(s/2),\coth(s/2)\big)\\ = \tau'_{-\sinh(s/2)\cosh(s/2)}\tau_{\tanh(s/2)}(0,\infty) = \tau'_{-\sinh(s)/2}\tau_{\tanh(s/2)}(0,\infty)
\end{multline}

\subsubsection{Super-additivity of the inner distance}

The super-additivity of the inner distance can certainly be shown directly. Moreover notice that, setting $I_r\equiv (-r,r)$, we have
\[
\ell(I_R,I_r) = \log\frac{R}{r}, \quad R>r>0\ ,
\]
and so in this case we have the additive property
\[
\ell(I_{r_1},I_{r_3}) = \ell(I_{r_1},I_{r_2})+\ell(I_{r_2},I_{r_3}),\quad r_1>r_2>r_3>0\ .
\]
Yet, it is instructive to give a functional analytic proof of the super-additivity.
For each $k\in\mathbb N$ there exists exactly one irreducible, positive 
energy, unitary representation (up to unitary equivalence) of $G$ with lowest
weight $k$, that we denote by $U^{(k)}$ with corresponding conformal
Hamiltonian $L_0^{(k)}$.  Then each eigenvalue of $L_0^{(k)}$ has
multiplicity one. 
Now
\[
t_{k,\tilde I, I}\equiv ||T^{U^{(k)}}_{\tilde I, I}||_1 =\Tr(e^{- s L_0^{(k)}})= \sum_{n=k}^{\infty} 
e^{-sn}
=\frac{e^{-sk}}{1-e^{-s}}
\]
$s=\ell(\tilde I, I)$.
We thus have
\begin{equation}\label{sub}
    t_{\tilde I, I}\equiv\lim_{k\to\infty}\sqrt[k]{t_{k,\tilde I, I}} =e^{-\ell(\tilde I, I)}
\end{equation}
Then $\ell$ is super-additive, namely
$t=e^{-\ell}$ is sub-multiplicative:
\[
t_{I_3, I_1}\leq t_{I_3, I_2}t_{I_2, I_1}
\]
with $I_3\Supset I_2\Supset I_1$.
This follows at once by eq. \eqref{sub} because $T^{U^{(k)}}_{I_3, I_1}=T^{U^{(k)}}_{I_2, I_1}T ^{U^{(k)}}_{I_3, I_2}$ and the nuclear norm is sub-multiplicative.

\subsection{Comments on an inequality in \cite{BDL2}}
\label{bdl2}

Let $\A$ be a Poincar\'e covariant net local von Neumann algebra as in Sect. \ref{Mink}. The following inequality is proved in \cite{BDL2}:
\begin{equation}\label{kd0}
||e^{-\tan(2\pi\l) d_{\O} H}\Delta_{\O}^{-\l}||\leq 2\ , \qquad 0<\l<1/4\ .
\end{equation}
Here $H$ is the time-translation generator, $\O$ is a double cone of $\mathbb R^{d+1}$ whose axis is contained in the time axis, $\D_{\O}$ is the vacuum modular operator associated with $\A(\O)$, and $d_{\O}$ is the time extension of $\O$.

We comment here more that the left hand side is bounded by 1, namely
\begin{equation}\label{kd}
||e^{-\tan(2\pi\l) d_{\O} H}\Delta_{\O}^{-\l}||\leq 1\ , \qquad 0<\l<1/4\ .
\end{equation}
Indeed we may consider the $n$-tensor product net $\A\otimes\cdots\otimes\A$ and apply eq. \eqref{kd0} to it. Then
\[
||e^{-\tan(2\pi\l) d_{\O} H}\Delta_{\O}^{-\l}||^n\leq 2
\]
that gives \eqref{kd0} as $n$ is arbitrary.

If one reads \eqref{kd} in the one-particle Hilbert space of a free field, one finds an inequality that only refers to representations of the Poincar\'e group.  In other words, if $U$ is a positive energy, unitary representation of
the Poincar\'e group, then equation \eqref{kd} holds true, where $\D_{\O}$ is the modular operator of the real Hilbert space associated with $\O$ (intersection of all real Hilbert subspaces associated with wedges that contain $\O$, see \cite{BGL3}).

Note that letting $\l\to 0^+$ in the inequality in the remark following \cite[Lemma 3.6]{BDL2} we get the operator inequality
\[
K_{\O} \leq d_{\O}\cdot H
\]
where $K_{\O} = -\frac{1}{2\pi}\log\D_{\O}$.
\subsection{Proof of Theorem \ref{magic'} in the general case}
\label{Am}

We end now the proof of Theorem \ref{magic'} for general representations of the universal cover $G$ of $SL(2,\mathbb R)$. This is necessary for further applications on nets on the cover of $S^1$. We begin to restate a result contained in \cite{GLW}.

\begin{Lemma}\label{GLW} {\rm \cite[Prop. 2.2]{GLW}}.
Let $U=U^{(1)}$ be the irreducible unitary representation of $G$ with lowest weight 1 on a Hilbert space $\H$ and denote by $K_1$, $H$ and $L_0$ the associated generators of the $\Iu$-dilations, translations and rotations. There exists a irreducible unitary representation $U^{(\alpha)}$ of $G$ with lowest weight $\alpha$ on $\H$, for all $\alpha\geq 1$ such that
\begin{align*}
K_1^{(\alpha)} &= K_1\\
H^{(\alpha)} &= H\\
L_0^{(\alpha)} &= L_0 - \lambda H^{-1},\qquad \lambda = \alpha(\alpha -1)/2\ .
\end{align*}
The above operator identities hold on a common core for all $K_1^{(\alpha)}$, $H^{(\alpha)}$ and $L_0^{(\alpha)}$, the corresponding generators for $U^{(\alpha)}$, jointly for all $\alpha\geq 1$.
\end{Lemma}
We recall that the construction in the \cite[Prop. 2.2]{GLW} was set up with $U^{(\alpha)}$ extending the Schroedinger representation for $K_1-\log H$ on $\H=L^2(\mathbb R)$. It is implicit in that proof  that $S(\mathbb R)$ is a joint core as stated in the above lemma.

Now we have $2L_0 = H + H'$, therefore ${H'}^{(\alpha)} = H' - 2 \lambda H^{-1}$. 
With $K_2\equiv K_{\Ir}$ we have $2K_2 = H - H'$ therefore
\begin{equation}\label{ka}
K_2^{(\alpha)} = K_2 +  \lambda H^{-1}
\end{equation}

\smallskip

\noindent
{\it End of proof of Theorem \ref{magic'}}.
Note first that in the $SL(2,\mathbb R)$ case we have (e.g. by Cor. \ref{hol})
\[
||\D_1^{iz}\D_2^{-is}\D_1^{-iz}||\leq 1,\quad z\in S(0,-1/2),s\in\mathbb R
\]
and therefore, by what already proved,
\begin{equation}\label{b1}
||e^{2\pi i s \big(\cosh(2\pi z)K_2 - \sinh(2\pi z)L_0\big)}||\leq 1,
\end{equation}
for all $z\in S(0,-1/2)$ and $s\in\mathbb R$.

{\it Part 3.}
In this part we assume that equation \eqref{b1} holds true. Let $\eta\in\H$ and choose a sequence of vectors $\eta_n\in\cal D$ norm converging to to $\eta$. Then $W_s(z)\eta_n \to W_s(z)\eta$ uniformly in $z$ showing that $W_s(\cdot)$ is holomorphic in $S(0,-1/2)\cap B_r$.

Because of the identity \eqref{BCH} we infer that the map
\[
t\mapsto \D_1^{it}\D_2^{-is}\D_1^{-it}
\]
has an analytic continuation in $S(0,-1/2)\cap B_r$. According to Thm. 2.3 of Araki and Zsido \cite{AZ} the value of this analytic continuation at $t = -1/4$ is the closure of $\D_1^{1/4}\D_2^{-is}\D_1^{-1/4}$. This must coincide with $W_s(t)_{t=-i/4}$, namely eq. \eqref{magic'} holds true.

We have therefore proved the theorem by assuming eq. \eqref{b1}.

{\it Part 4.} In this part we assume that $U(2\pi)$, the $2\pi$-rotation in the representation $U$ is a scalar and $U(2\pi)= e^{i2\pi\vartheta}$ with $\vartheta$ rational. Thus $U$ is a representation of a finite cover of $SL(2,\mathbb R)$.

Then there exists $n\in\mathbb N$ such that the $n$-tensor product $\tilde U\equiv U\otimes U\otimes\cdots \otimes U$ is a representation of $SL(2,\mathbb R)$. By part 1 equation \eqref{b1} holds in the representation $\tilde U$, namely $||W_s(z)\otimes\cdots\otimes W_s(z)||\leq 1$. Therefore $||W_s(z)||\leq 1$, namely equation \eqref{b1} holds in the representation $U$. By part 3 we conclude that eq. \eqref{magic'} is true in the representation $U$.

{\it Part 5.} In this part we assume that $U=U^{(\alpha_0)}$ is the irreducible representation of $G$ with lowest weight $\alpha_0$ irrational. Therefore $U(2\pi)$ is a scalar. We assume first that $\alpha_0>1$ so we may be in the setting of Lemma \ref{GLW} and equation \eqref{ka} holds too. Let $\eta\in\cal D$ be a unit vector and consider the function
\begin{align*}
F(s,z,\l) &\equiv 
e^{2\pi i s \big(\cosh(2\pi z)K^{\alpha}_2 - \sinh(2\pi z)L^{\alpha}_0\big)} \eta\\
&= e^{2\pi i s \big(\cosh(2\pi z)K_2 - \sinh(2\pi z)L_0
+\l^{-1}(\cosh(2\pi z)+\sinh(2\pi z))H^{-1} \big)}\eta\\
&= e^{2\pi i s \big(\cosh(2\pi z)K_2 - \sinh(2\pi z)L_0
+\l^{-1}(\exp(2\pi z))H^{-1} \big)}\eta
\end{align*}
$\l=\alpha(\alpha - 1)$. Note that $F$ is defined $|s|\leq s_0$, $z\in S(0,1/2)$ and $\alpha$ the given $\alpha_0$. By part 4, $F$ is also defined if $\alpha \geq 1$ is rational (and all $s\in\mathbb R$ and  $z\in S(0,1/2)$) and in this case $||F(s,z,\l)||\leq 1$.

As the map 
\[
s,z,\l\in [-s_0,s_0]\times S(0,\frac12)\times \mathbb R_+\mapsto s\l^{-1}e^{2\pi z}
\]
is open (by the holomorphic open mapping theorem), it follows that if $\l$ is sufficiently close to $\l_0=\alpha_0(\alpha_0 - 1)$, there exists $s'$ and $z'$ such that $F(s',z',\l)=F(s,z,\l_0)$. Taking $\alpha\in \mathbb Q$ we conclude that $||F(s,z,\l_0)||\leq 1$. 

Therefore $||W_s(z)||\leq 1$. By part 3 we have proved the lemma for $U^{(\alpha_0)}$ with $\alpha_0>1$.  

Concerning the case $0<\alpha_0 <1$, by consider $U^{(\alpha_0)}\otimes U^{(2-\alpha_0)}$ we conclude as in part 4 that $||W_s(z)||\leq 1$ for $U^{(\alpha_0)}$, thus the theorem holds in this case by part 3.

{\it Part 6.} We may now finish our proof. Indeed if $U$ is any unitary, positive energy representation of $G$, then $U$ has a direct integral decomposition
\[
U = \int^{\oplus}_{\mathbb R^{+}}N(\alpha)U^{(\alpha)}{\rm d}\alpha
\]
for some multiplicity function $N$. As the statement holds for $U^{(\alpha)}$ it then holds for $U$.
\endproof
As a corollary, setting $t=-i/2$ in Theorem \ref{magic'} we see that a formula in \cite{GL} concerning a class of unitary representations of $SL(2,\mathbb R)$ also holds for all unitary, positive energy representations of $G$:
\begin{Cor}
If $U$ is a unitary, positive energy representation of $G$ and $s>0$, the associated operators as in Th. \ref{magic'} satisfy
\begin{equation}\label{GL}
\D_1^{1/2}\D_2^{-is}\D_1^{-1/2}\xi = \D_2^{is}\xi
\end{equation}
for $\xi$ in a core for $\D_1^{-1/2}$.
\end{Cor}

\section{Concluding comments}
There are two natural problems left open in this paper. The first is whether $L^2$-nuclearity holds for double cone inclusion associated with the free, scalar, massive field. An answer to this question would clarify the meaning of the $L^2$-nuclearity condition, in particular whether it is tight up with conformal invariance and/or spacetime with a space compactification. 

The second problem is whether BW-nuclearity in chiral conformal QFT implies the trace class condition. In other words one would infer from BW-nuclearity condition conformal nuclearity with a uniformity in interval length.

There is a natural continuation of our work concerning a discussion of the nuclearity condition for chiral conformal nets in non-vacuum sectors, but this goes beyond our aim in this article.

\bigskip

\noindent
{\bf Acknowledgements.} We thank H. Bostelmann for a helpful comment on the material in Section 8.2.


{\footnotesize \begin{thebibliography}{99}

\bibitem{AZ} H. Araki \& L. Zsido: {\it Extension of the structure theorem of Borchers and its application to half-sided modular inclusions}, Rev. Math. Phys. {\bf 17} (2005), 491-543.

\bibitem{B} H.J. Borchers, {\it The CPT theorem in two-dimensional
theories of local observables}, Commun. Math Phys. {\bf 143} (1992), 315-332.
 
\bibitem{BB} H.J. Borchers \& D. Buchholz, {\it Global properties of vacuum states in de Sitter space}.
 Ann. Inst. H. Poincar\'e Phys. Th\'eor. {\bf 70}  (1999), 23-40. 

\bibitem{BGL1} R. Brunetti, D. Guido \& R. Longo, 
{\it Modular structure and duality in conformal quantum field theory}, 
Commun. Math. Phys. {\bf 156}, 201-219, (1993).

\bibitem{BGL3} R. Brunetti, D. Guido \& R. Longo, 
{\it Modular localization and Wigner particles}, Rev. Math. Phys. {\bf 14}, 759-785 (2002) 

\bibitem{BW} D. Buchholz \& E. Wichmann,
{\it Causal independence and the energy-level density of states in local quantum field theory}, Commun. Math. Phys. {\bf 106}, 321 (1986).

\bibitem{BDL1} D. Buchholz, C. D'Antoni \& R. Longo: {\it Nuclear maps and modular structures. I. General properties}, J. Funct. Anal. {\bf 88}, 233-250 (1990).

\bibitem{BDL2} D. Buchholz, C. D'Antoni \& R. Longo: {\it Nuclear maps
and modular structures II: application to quantum field theory},
Commun. Math. Phys. {\bf 129}, 115-138 (1990).

\bibitem{BL} D. Buchholz \&  G. Lechner: {\it Modular nuclearity and localization}                                                 
Ann. H. Poincar\'e   {\bf 5},  1065 - 1080  (2004).                           

\bibitem{BJa} D. Buchholz \& Jacobi: {\it On the nuclearity condition for massless fields}, Lett. Math. Phys.  {\bf 121}  (1989),  255-270.

\bibitem{BJ} D. Buchholz \& P. Junglas: {\it On the existence of equilibrium states in local quantum field  theory},  Commun. Math. Phys.  {\bf 13}  (1987),  313-323.

\bibitem{BY} D. Buchholz \& J. Yngvason, {\it Generalized nuclearity conditions and the split property in quantum field theory}, Lett. Math. Phys. {\bf 23} (1991), 159-167.

\bibitem{DLR} C. D'Antoni, R. Longo \& F. Radulescu: {\it Conformal nets, maximal temperature and models from free probability}, J. Operator Theory {\bf 45}, 195-208 (2001).

\bibitem{DDLF} C. D'Antoni,  S. Doplicher, K. Fredenhagen \& R. Longo: {\it  Convergence of local charges and continuity properties of  W$^*$-inclusions}, Commun. Math. Phys. {\bf 110} (1987),      325-348.

\bibitem{DF} C. D'Antoni \& K. Fredenhagen: {\it  Charges in space-like cones},  Commun. Math. Phys. 94 (1984), 537-544.

\bibitem{DL} S. Doplicher \& R. Longo: {\it  Standard and split inclusions of von Neumann algebras}, Invent. Math. 75 (1984), 493-536.

\bibitem{GL} D. Guido \& R. Longo, {\it An algebraic spin and statistics theorem}, Commun. Math. Phys. {\bf 172} (1995), 517-533.

\bibitem{GLW} D. Guido, R. Longo \& H.-W. Wiesbrock, 
{\it Extensions of conformal nets and superselection structures}, Commun. Math. Phys. {\bf 192}, 217-244 (1998).

\bibitem{H} R. Haag, ``Local Quantum Physics", Springer-Verlag 1996.

\bibitem{HL} P.D. Hislop \& R. Longo, {\it Modular structure of the local algebras associated with the free massless scalar field theory}, Commun. Math. Phys. {\bf 84} (1982), 71-85.

\bibitem{KL} Y. Kawahigashi \& R. Longo: {\it Noncommutative spectral invariants and black hole entropy}, Commun. Math. Phys. {\bf 257} (2005), 193-225.

\bibitem{K} A.A. Kirillov, ``Elements of the Theory of Representations'', Springer-Verlag, New York, 1976.

M. Sugiura, ``Unitary Representations and Harmonic Analysis. An Introduction'', 
North-Holland, Amsterdam, Tokyo 1990.

\bibitem{L}  R. Longo, {\it Notes for a quantum index theorem}, Commun. Math. Phys. {\bf 222} (2001), 45-96.

\bibitem{NW} E. Nelson: {\it Analytic vectors}, Ann. Math. {\bf 70} (1959), 572-615.

R. Goodman: {\it Analytic and entire vectors for representations of Lie groups}, Trans. Amer. Mat. Soc. {\bf 143} (1969), 55-76.

\bibitem{R} M.B. Ruskai: {\it Inequalities for traces on von Neumann algebras}, Commun. Math. Phys. {\bf 26}, 280-289 (1972). 

\bibitem{S} B. Schroer: {\it Two-dimensional models as testing ground for
principles and concepts of local quantum physics}, hep-th/0504206

\bibitem{SZ}S. Stratila \& L. Zsido, ``Lectures on von Neumann Algebras'', Abacus Press, Tunbridge Wells, Kent, England, 1979. 

\bibitem{W} H.-W. Wiesbrock: {\it Half-sided modular inclusions of von Neumann algebras}, Commun. Math. Phys.  {\bf 157}  (1993), 83--92.

\end{thebibliography}}
\end{document}